\documentclass[12pt]{article}

\usepackage{authblk}

\usepackage{xcolor}

\usepackage{graphicx} 
\usepackage[english]{babel}
\usepackage{bm,bbm,amsmath,amssymb,graphicx}
\usepackage{url}
\usepackage[title]{appendix}
\usepackage{algpseudocode}
\usepackage{algorithm}
\usepackage[colorlinks = true]{hyperref}
\usepackage{graphicx,subcaption}
\usepackage{verbatim}
\usepackage{bm}
\usepackage[numbers]{natbib}
\usepackage[margin=1in]{geometry}
\usepackage{ dsfont }
\numberwithin{equation}{section}

\def \RRR {\mathbb{R}}
\def \a {{\bm{a}}}

\def \b {\bm{b}}

\def \w {\bm{w}}

\def \X {\mathbf{X}}

\def \Y {\mathbf{Y}}

\def \bPhi {\boldsymbol{\Phi}}

\usepackage{ mathrsfs }
\newcommand{\col}[1]{#1_{(\cdot,j)}}
\newcommand{\row}[1]{#1_{(i,\cdot)}}
\newcommand{\est}[1]{#1^t}
\newcommand{\estprev}[1]{#1^{t-1}}
\newcommand{\bP}{{\mathds{P}}}
\def \bPhi 		{\bm{\Phi}}
\def \X				{\bm{X}}
\def \Y				{\bm{Y}}
\def \jS           {\mathscr{S}}	
\def \w			{\bm{w}}

\begin{document}

\title{Sparse Randomized Kaczmarz for Support Recovery of Jointly Sparse Corrupted Multiple Measurement Vectors}
\author[1]{Natalie Durgin}
\author[2]{Rachel Grotheer}
\author[3]{Chenxi Huang}
\author[4]{Shuang Li}
\author[5]{Anna Ma}
\author[6]{Deanna Needell}
\author[7]{Jing Qin}

\affil[1]{Spiceworks} 
\affil[2]{Goucher College}
\affil[3]{Yale University}
\affil[4]{Colorado School of Mines}
\affil[5]{Claremont Graduate University}
\affil[6]{University of California, Los Angeles}
\affil[7]{Montana State University}

\maketitle

\abstract{While single measurement vector (SMV) models have been widely studied in signal processing, there is a surging interest in addressing the multiple measurement vectors (MMV) problem. In the MMV setting, more than one measurement vector is available and the multiple signals to be recovered share some commonalities such as a common support. Applications in which MMV is a naturally occurring phenomenon include online streaming, medical imaging, and video recovery. This work presents a stochastic iterative algorithm for the support recovery of jointly sparse corrupted MMV. We present a variant of the Sparse Randomized Kaczmarz algorithm for corrupted MMV and compare our proposed method with an existing Kaczmarz type algorithm for MMV problems. We also showcase the usefulness of our approach in the online (streaming) setting and provide empirical evidence that suggests the robustness of the proposed method to the distribution of the corruption and the number of corruptions occurring. }

\section{Introduction}
In recent years, there has been a drastic increase in the amount of available data. This so-called ``data deluge'' has created a demand for fast, iterative algorithms that can be used to process large-scale data.  Stochastic iterative algorithms, such as the Randomized Kaczmarz Algorithm or Stochastic Gradient Descent, have become an increasingly popular option for processing large-scale data \cite{bottou2010large, Kac37:Angenaeherte-Aufloesung}. These methods recover signals $\X \in \mathbb{R}^n$ given a vector of measurements  $\Y \in \mathbb{R}^{m}$ and a measurement matrix $\bPhi \in \mathbb{R}^{m \times n}$ such that:
\begin{equation}
\Y = \bPhi \X, 
\label{eq:smv}
\end{equation} 
without accessing the full measurement matrix in a single iteration. We refer to \eqref{eq:smv} as a Single Measurement Vector (SMV) model. 
In the Multiple Measurement Vector (MMV) setting, one may have thousands of measurement vectors $\col{\Y}$ pouring in overtime. Each measurement vector $\col{\Y}$ corresponds to a signal $\col{\X}$ where signals typically share a common property such as sparsity, smoothness, etc. For simplicity, let $\Y = [\Y_{(\cdot,1)} \cdots \Y_{(\cdot,J)}] \in \RRR^{m\times J}$ and  $\X = [\X_{(\cdot,1)} \cdots \X_{(\cdot,J)}] \in \RRR^{n\times J}$. Since high-dimensional data is typically sparse in nature, a commonality of particular interest is \emph{joint sparsity}, or when the support of all signals are the same. In particular, the support of a vector $\textbf{v}$ is defined to be the set of indexes for which $\textbf{v}$ is nonzero i.e., $supp(\textbf{v}) = \{ i : \textbf{v}_i \neq 0 \}$.

Many algorithms have been developed for the MMV setting, especially in applications such as line spectral estimation \cite{li2016off, yang2014exact} and modal analysis \cite{li2017atomic}. In particular, the authors in these works extend the previous SMV-based algorithms as well as theoretical analysis in \cite{candes2014towards, tang2013compressed,heckel2017generalized} to the MMV case. The theoretical bound in \cite{li2017atomic} also indicates that MMV settings could make compressed signal recovery much easier than in the SMV setting. In particular, the number of measurements needed for perfect recovery in each signal decreases as the number of signals increases reducing the sample complexity per signal.

As a motivating example, consider diffuse optical tomography (DOT) where the goal is to find small areas of high contrast corresponding to the location of cancerous cells \cite{arridgereview_new}. 
Since cancerous cells have a much larger absorption coefficient than healthy cells, the two-dimensional medical image 
can be interpreted as a sparse signal where each entry of the signal represents the absorption coefficient of a given pixel and the nonzero entries correspond to tumor locations.
In a hyperspectral DOT setting, hundreds of different wavelengths are used to acquire a variety of images of the same tissue, allowing practitioners to obtain a more accurate location of tumors~\cite{larusson}. 
The result of the hyperspectral imaging process is a jointly sparse MMV, where each wavelength produces a different image (or signal), and the joint support across all images represents the locations of cancerous cells.

Signals may share support but it is improbable for them to be perfectly accurate. 
Since sensing mechanisms are not impervious to error, signals can contain corruptions. 
Other sources of corruption in signal processing include spikes in power supply, defective hardware, and adversarial agents \cite{laska2009exact}. 
Going back to the hyperspectral imaging example, ``corruptions" in each signal may be caused by noncancerous cells that absorb more light at a given wavelength than their neighbors. For example, if a cell contains an anomalous amount of melanin, then it absorbs more light at shorter wavelengths in the visible spectrum (i.e., violet or blue light) compared to a typical noncancerous cell~\cite{lufei, fodor2011}. This produces a large nonzero absorption coefficient in the location of a healthy cell, i.e., a corruption. These corrupt entries erroneously indicate the presence of cancerous cells in a location with healthy cells. 

Corruptions cause support recovery algorithms such as the MMV Sparse Randomized Kaczmarz (MMV-SRK) algorithm, which we describe in detail in Section \ref{sec:related}, to fail due to the algorithmic dependence on the row norms of the signal approximation to estimate the support \cite{aggarwal2014extension}. Thus, large corruptions in a signal with comparatively small entries may erroneously be included in the support estimate given by these algorithms. In the corrupt MMV setting, the availability of multiple measurement vectors becomes vital to the estimate of the true support. Clearly, if only a single measurement vector is available, there would be no way to distinguish a corrupt nonzero entry without any additional assumptions on the signal or corruption. Corrupt measurement signals have been studied in the context of the SMV model. In \cite{studer2012recovery} and \cite{laska2009exact}, additive noise in the measurement scheme is assumed to be sparse. Both works focus on the compressive sensing setting where $m \ll n$. 

The primary objective of this work is to design an algorithm for recovering the support of jointly sparse, corrupt signals from large-scale MMV. We propose a new online algorithm called Sparse Randomized Kaczmarz for Corrupted MMV (cMMV-SRK) for support recovery. 
Note that the proposed algorithm can recover the signals very well, but we mainly focus on support recovery in this work.
Our experiments show that the proposed algorithm outperforms the previously proposed Kaczmarz type algorithm in recovering the joint support from MMV when the signals are corrupted.

\subsection{Problem Formulation}
The mathematical formulation of the problem can be stated as follows. Suppose one is given a set of linear measurements $\col{\Y} \in \mathbb{R}^{m}$ and a measurement matrix $\bPhi \in \mathbb{R}^{m \times n}$ such that: 
\begin{equation}
\col{\Y} = \bPhi \col{\X} \quad\quad\quad \text{ for } j = 1, \dots J,
\end{equation}
with $\col{\X} \in \mathbb{R}^{n}$. We assume that the data is large-scale, meaning we cannot access all of $\bPhi$ at once ($m$ and/or $n$ are too large) and must only operate on one row of $\bPhi$ at a time.  
We allow the system to be overdetermined ($m \gg n$) or underdetermined ($m \ll n$) and assume $\col{\X}$'s are jointly sparse such that $supp(\col{\X}) = \jS$ and $|\jS| = k$. 
For an $n$-dimensional vector $\col{\X}$, let $\col{\X}\rvert_{s}$ return $\col{\X}$ with zeros in the $n - s$ smallest (in magnitude) entries. We also assume that each column of $\X$ contains one or more \emph{corruptions}. In other words, instead of $ supp(\col{\X})  \subset  \jS$, the joint support set, the support of $\col{\X}$ is:
\begin{equation*}
supp(\col{\X}) = \jS \cup \mathcal{C}_j, \quad\quad\quad \mathcal{C}_j \subset \{ 1, \dots, n \}, \label{eq:corrupt}
\end{equation*}
where $\mathcal{C}_j$ is the ``corrupt index set" and $\mathcal{C}_j$ are not necessarily the same for every $j$. In this work, our goal is to recover the joint support $\jS$ from the given linear measurements $\Y$. 

The remainder of this manuscript is organized in the following way. Section \ref{sec:related} discusses the Sparse Randomized Kaczmarz method and
  the MMV-SRK algorithm. Section \ref{sec:main} 
  provides a discussion on how corruptions can negatively impact the performance of MMV-SRK. Section \ref{sec:main} also presents our method, cMMV-SRK, a variant of SRK which works in the corrupted signal setting. Numerical experiments using this method are presented in Section \ref{sec:exp} and we conclude with a summary of our contributions and future directions in Section \ref{sec:conclu}.

\section{Related and Existing Work}
\label{sec:related}

\subsection{Sparse Randomized Kaczmarz}
In this work, we utilize the Sparse Randomized Kaczmarz algorithm to recover the support of each column of $\X$. The original Kaczmarz algorithm was first introduced in the early 1900s by Kaczmarz himself and was revitalized as the Algebraic Reconstruction Technique in the early 1980s \cite{Kac37:Angenaeherte-Aufloesung, GBH70:Algebraic-Reconstruction}. The Randomized Kaczmarz Algorithm (RK) used here was first introduced by Strohmer and Vershynin and enjoys an expected linear convergence rate to the solution of a consistent linear system \cite{SV09:Comments-Randomized}. The Sparse Randomized Kaczmarz (SRK) algorithm is another variant designed specifically for overdetermined systems with sparse solutions. SRK has also been empirically shown to solve underdetermined systems with sparse solutions as well \cite{MY13}.

Algorithm \ref{alg:srk} outlines the SRK algorithm. Note that ties are broken lexicographically in Step 5 of Algorithm \ref{alg:srk} and all algorithms presented in this work. The estimated support size $\hat{k}$ is a parameter of the algorithm and is typically chosen to be larger than the true support size $k$. In this variant, the algorithm runs for a specified number of iterations (up to $\tau$). However, any stopping criteria one would use for an iterative algorithm e.g. terminating after the residual meets a certain criteria, after the updates become small, etc. can be used. Algorithm \ref{alg:srk} also differs from the original presented algorithm by \cite{MY13} in that at every iteration the support estimate has size $\hat{k}$ instead of starting with $n$ and shrinking the size to $\hat{k}$. We find that these modifications do not significantly affect the behavior of SRK.

\begin{algorithm}[t!]
\caption{Sparse Randomized Kaczmarz}\label{alg:srk}
\begin{algorithmic}[1]
\Procedure{SRK}{$\bPhi \in \mathbb{R}^{m \times n} $, $\Y \in \mathbb{R}^{m}$, $\hat{k} $, $\tau$}
\State Initialize $\X^0 = \textbf{0}^{n \times 1}$
\For{$ t = 1, ..., \tau$}
\State Choose row $\row{\bPhi}$  with probability  $\frac{\|\row{\bPhi}\|_2^2}{\|\bPhi\|_F^2} $
\State Set support estimate $\est{\jS} = supp(\estprev{\X} \rvert_{\hat{k}})$
	\State Set row weights \Comment{$w \in \mathbb{R}^{n}$} \[w_l = \left\{
  \begin{array}{lr}
    1 &: l \in \est{\jS} \\
    \frac{1}{\sqrt{t}} &: l \in {\jS^t}^c
  \end{array}
\right.
\]    \Comment{${\jS^t}^c$ is the complement set of $\est{\jS}$} 
	\State Set $\a =  \w \cdot \row{\bPhi}$ \Comment{$\a \in \mathbb{R}^{n}$ is the weighted row of $\bPhi$}
	\State Update $\est{\X} = \estprev{\X} + \frac{\Y_i - \a\estprev{\X}   }{\|\ \a\|^2_2} \a^T$ 
\EndFor
\State \textbf{return} $\X^\tau$
\EndProcedure
\end{algorithmic}
\end{algorithm}

Algorithm \ref{alg:srk} has been shown empirically to find the solution to overdetermined, consistent (i.e., a solution exists) linear systems but there are no theoretical results supporting this. One can make a few observations about the behavior of SRK for support recovery. Concerning the support size estimate $\hat{k}$, it is clear that if $\hat{k} < k$ then the probability that the true support is contained in the support of the approximation is 0, i.e., $\bP \left( \jS \subset supp(\X^\tau) \right) = 0$.
Additionally, if $\hat{k} = n$, then $\bP \left( \jS \subset supp(\X^\tau) \right)=1$. In regards to the choice of weighting, as $t \rightarrow \infty$, $\frac{1}{\sqrt{t}} \rightarrow 0$ so that row elements inside the support estimate contribute mostly to the approximation. If one has a weighting function that decreases too rapidly, the true support may not be captured in $\est{\jS}$ causing the algorithm to fail. If the weighting function decreases too slowly, then the algorithm will converge more slowly.

Although Algorithm~\ref{alg:srk} and the following algorithms require the Frobenious norm of the matrix, $\|\bPhi\|_F^2$, for row selection, practically speaking row selections can be done uniformly at random to avoid using the full measurement matrix in a single iteration. Indeed, it is advantageous to select the rows at random to avoid introducing bias from rows with larger norms. 

\subsection{SRK for MMV}
Here we present a previous SRK-based approach to the MMV setting proposed by \cite{aggarwal2014extension}. Because we are assuming joint sparsity in the MMV model, the estimated support of a signal reveals information about the support of all signals. The authors of \cite{aggarwal2014extension} present Algorithm \ref{alg:mmvsrk} to leverage this idea. There are a few key aspects to note about this version of the SRK algorithm. First, the algorithm is running one iteration of SRK for every signal in the MMV model then updating the support estimate based on the row norms of the estimate $\est{\X}$. Due to this, the algorithm does not lend itself well to being extended for an online variant which only receives a small number (possibly 1) of signals at a time.  
Second, the algorithm uses the same selected row for each signal. It has been well observed that a random selection scheme reduces the possibility of a poor choice of row ordering and it may be advantageous to allow each signal to be projected onto a different randomly selected row \cite{HS78:Angles-Null,HM93:Algebraic-Reconstruction}

\begin{algorithm}[h!]
\caption{Sparse Randomized Kaczmarz for MMV}\label{alg:mmvsrk}
\begin{algorithmic}[1]
\Procedure{MMV-SRK}{$\bPhi\in \mathbb{R}^{m \times n}$, $\Y \in \mathbb{R}^{m \times J}$, $\hat{k}$, $\tau$}
\State Initialize $\X^0 = \textbf{0}^{n \times J}$ 
\For{$ t = 1, \dots,  \tau$}
\State Choose row $\row{\bPhi}$  with probability  $\frac{\| \row{\bPhi} \|_2^2}{\|\bPhi\|_F^2} $
\State Set support estimate $\est{\jS}$: $\hat{k}$ indices with largest row norm of $\estprev{\X}$ 
	\State Set row weights \Comment{$w \in \mathbb{R}^{n}$} \[w_l = \left\{
  \begin{array}{lr}
    1 &: l \in \est{\jS} \\
    \frac{1}{\sqrt{t}} &: l \in  {\jS^t}^c
  \end{array}
\right.
\]   \Comment{ ${\jS^t}^c$ is the complement set of $\est{\jS}$} 
	\State Set $\a = \w \cdot \row{\bPhi}$ \Comment{$\a$ is the weighted row of $\bPhi$}
	\For{ $j = 1, \dots, J$}
	\State Update $\est{\col{\X}} = \estprev{\col{\X}} + \frac{\Y_{(i,j)} - a\estprev{\col{\X}}}{\| \a\|^2_2} \a^T$ 
	\EndFor
	\State Update $\est{\X} = [\est{\X_{(\cdot, 1)}} | \dots | \est{\X_{(\cdot, J)}}] $
\EndFor
\State \textbf{return} $\est{\X}$
\EndProcedure
\end{algorithmic}
\end{algorithm}

\section{Main Results}

\label{sec:main}
\subsection{Corrupted MMV} 
\label{sec:corrupt}
To review, we are interested in constructing an algorithm that recovers the support of jointly sparse corrupted high-dimensional MMV, that is, where we can only access one row of the measurement matrix $\bPhi$ at a time. 
To this end, we propose Algorithm~\ref{alg:c_mmvsrk}, which we refer to as cMMV-SRK. 
We first note that the base of this algorithm is the SRK algorithm (Algorithm~\ref{alg:srk}), which is an effective algorithm for large-scale problems due to its low memory footprint, requiring only one row of the measurement matrix to be used at a time.  
cMMV-SRK has been adapted to the MMV case using the intuition 
that the individual signals give us information about the common support between all signals. We keep track of a bin or tally vector $\b$ that estimates the true support of the signals. 
In particular, we use the nonzeros in $\b$ to indicate the estimated joint support. 
This binning process allows the algorithm to be robust in the face of corruptions in the signal, as the corruptions will receive a low number of tallies compared to the entries in the true support 
because the corruptions occur in random positions for every signal.
Note that in the corrupted MMV case, we expect Algorithm \ref{alg:mmvsrk} to fail as the support estimate step relies on the $\ell_2$-norm of the rows to be large if an index is in the support and small otherwise. 
The corruptions may be so large that a single corruption in a row could lead to mis-identification of the corrupt entry being in the joint support. 

Finally, in Algorithm~\ref{alg:c_mmvsrk} we account for signals being processed one at a time, as they are in an online or ``streaming" setting. 

For each signal, let $\tilde{\tau}_j$ be the number of SRK projections performed on $\col{\X}$ and let $\tilde{\tau} = [\tilde{\tau}_1 \cdots \tilde{\tau}_J]$. In the online setting, one can imagine that the amount of time before the next signal is acquired may vary due to, for example, stalls in the measurement process.
The varying amount of time that the system has to process each signal is one of the major challenges of support recovery in the online setting.
In order to improve the joint support estimate when $\tilde{\tau}_j$ varies, we weight the binning based on $\tau = \max_t \tilde{\tau_t}$. In other words, we let $\b_q = \b_q +\frac{\tilde{\tau_j}}{\tau}$ where $\b_q$ is the $q$-th entry of $\b$ and $\tilde{\tau}_j$ is the maximum number of inner iterations of SRK for the $j^{th}$ signal. This reweighting scheme places a larger importance on support estimates which have had more time (iterations) to improve. In the online setting where $\tilde{\tau}_j$s is not known a priori, $\tau$ can be set manually. 
We adopt the following notation for cMMV-SRK: $\est{S_j}$ is the estimated support at the $t^{th}$ SRK iteration for $\col{\X}$, and $\est{\jS}$ denotes the joint support estimate.

\begin{algorithm}[ht!]
\caption{Sparse Randomized Kaczmarz for Corrupted MMV}\label{alg:c_mmvsrk} 
\begin{algorithmic}[1]
\Procedure{cMMV-SRK}{ $\bPhi\in \mathbb{R}^{m \times n}$, $\Y \in \mathbb{R}^{m \times J}$, $\hat{k}$, $\tilde{\tau}$} 
\State Initialize $\X^0= \textbf{0}^{n \times J},~\b = \textbf{0}^{n \times 1}$, $\tau = \max_j \tilde{\tau_j}$	
\For{ $j = 1, \dots, J$}
	\For{$ t = 1, \dots,  \tilde{\tau}_j$} 
	\State Choose row $\row{\bPhi}$  with probability  $\frac{\|\row{\bPhi}\|_2^2}{\|\bPhi\|_F^2} $
\State Set support estimate $\est{S_j} = supp(\estprev{\col{\X}} \rvert_{\hat{k}})$
	\State Set row weights \Comment{$w \in \mathbb{R}^{n}$} \[w_l = \left\{
  \begin{array}{lr}
    1 &: l \in \est{S} \\
    \frac{1}{\sqrt{t}}  &: l \in {S^t}^c
  \end{array}
\right.
\]   \Comment{${S^t}^c$ is the complement set of $\est{S}$} 
	\State Set $\a = \w \cdot \row{\bPhi} $  \Comment{$\a$ is the weighted row of $\bPhi$}
	\State Update $\est{\col{\X}} = \estprev{\col{\X}} + \frac{\Y_{(i,j)} - \a \estprev{\col{\X}}  }{\|\a\|^2_2} \a^T$  
	\EndFor
	\State If $q \in \est{S_j}$ then $\b_q = \b_q +\frac{\tilde{\tau}_j}{\tau}$
	\State Set initial support estimate for next signal $\jS^j = supp(\b \rvert_{\hat{k}})$
\EndFor
\State \textbf{return} Joint support estimate $\jS^J$
\EndProcedure
\end{algorithmic}
\end{algorithm}

If the number of inner iterations $\tilde{\tau}_j$ is large enough, the support estimate should be such that it contains the joint support (along with the corruption index). Because we are tallying the support estimate after every $\tau$ iterations, it is clear that the entries in the joint support will have an overwhelming number of tallies compared to all other entries. The experimental results in the next section support these claims and we leave the analytical study of these algorithms for future work.

\section{Experiments}
\label{sec:exp}
In this section, we compare Algorithm \ref{alg:mmvsrk} and Algorithm \ref{alg:c_mmvsrk} under a variety of settings, specifically comparing the robustness of each algorithm in the presence of corruptions in the signal.
To test this robustness, we vary the number of corrupt entries, the distribution from which the corruptions are drawn, the number of columns in $\X$, and the number of projection computations $\tilde{\tau}_j$ made for each signal. In what follows, we will refer to $\tau$ as the number of SRK iterations.
These experiments are summarized in Table \ref{tab:summary}. In all experiments, the results are averaged over 40 trials and the nonzero entries of $\X$ are drawn independently and identically distributed (i.i.d.) from $\mathcal{N}(0,1)$. Note that on the $x$-axis we plot the number of ``iterations" where a single iteration is defined by a projection. In other words, the $x$-axis represents every time Step 9 is performed in Algorithm \ref{alg:c_mmvsrk} and Algorithm \ref{alg:mmvsrk}.

\begin{table}[h!]
\centering
\caption{Summary of Experiment Parameters. This table provides a summary of experiment parameters and Figure references for each experiment.}
\label{tab:summary}
\begin{tabular}{|c|c|c|c|c|c|}
\hline
\textbf{Figures}       & \textbf{$\bPhi$ entries} & \textbf{Dist. of Corruptions} & \textbf{Num. of Corruptions} & \textbf{$\tilde{\tau}_j$} & \textbf{$J$} \\ \hline
Figure \ref{fig:f1a} & Gaussian                & $\mathcal{N}(7,1)$                   & 1                              & 40              & 300          \\ \hline
Figure \ref{fig:f1b} & Uniform                 & $\mathcal{N}(7,1)$                   & 1                              & 80              & 600          \\ \hline
Figure \ref{fig:f2a} & Gaussian                & $\mathcal{N}(0,1)$                   & 1                              & 40              & 300          \\ \hline
Figure \ref{fig:f2b} & Uniform                 & $\mathcal{N}(0,1)$                   & 1                              & 80              & 600          \\ \hline
Figure \ref{fig:f3} & Gaussian                & $\mathcal{N}(7,1)$                   & Varies                         & 50              & 300          \\ \hline
Figure \ref{fig:f4} & Gaussian                & $\mathcal{N}(7,1)$                   & Varies                         & Varies          & 800          \\ \hline
\end{tabular}
\end{table}

\begin{figure}[h!]
\centering
\begin{subfigure}{0.4\textwidth}
\includegraphics[width=\textwidth]{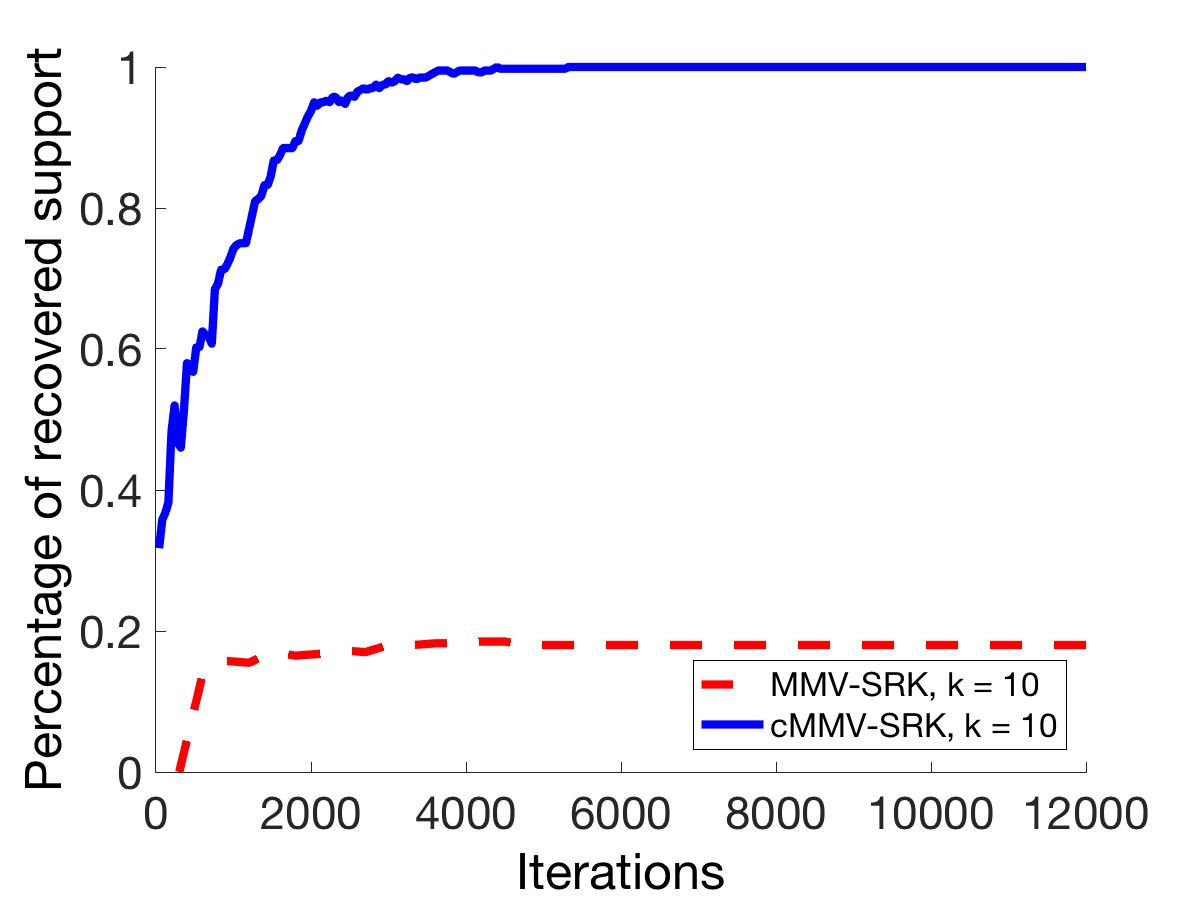}
\caption{$\bPhi \sim \mathcal{N}(0,1)$}
\label{fig:f1a}
\end{subfigure}
\begin{subfigure}{0.4\textwidth}
\includegraphics[width=\textwidth]{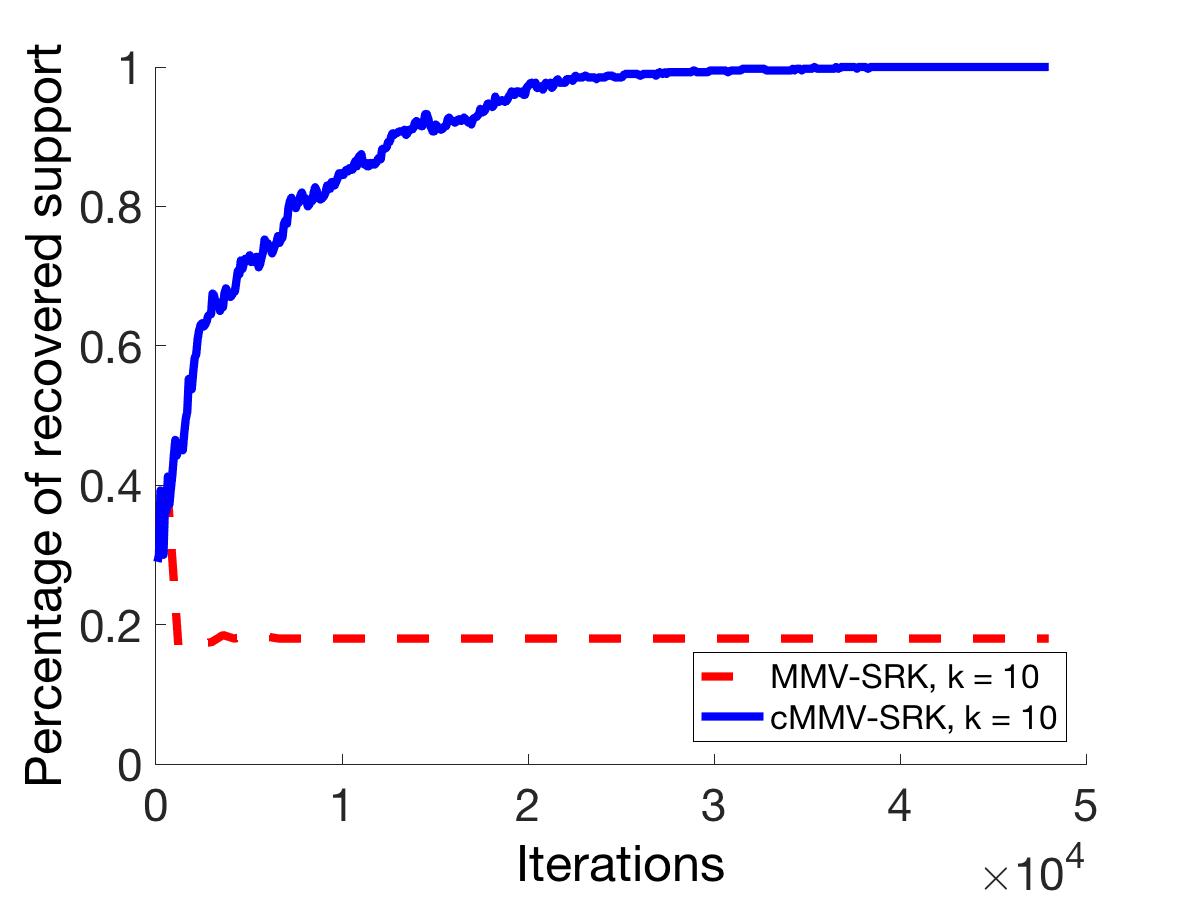}
\caption{$\bPhi \sim Unif([0,1])$}
\label{fig:f1b}
\end{subfigure}
\caption{Comparing SRK and MMV-SRK for support recovery when there is a single corrupt entry per signal whose magnitude is drawn from $\mathcal{N}(7,1)$.}
\label{fig:f1}
\end{figure}

Figure \ref{fig:f1} compares Algorithm \ref{alg:mmvsrk} and Algorithm \ref{alg:c_mmvsrk} with $m = 1000$, $n = 100$, and $k = 10$. The support size estimate is $\hat{k} = 1.5k$. To create $\X$, we uniformly at random select $k$ of $n$ indexes to be joint support $\jS$ and set $\Y = \bPhi \X$. The corrupt entries are drawn uniformly at random from $\{1,...N \} \setminus \jS$. To start off, each signal has one corrupt entry. We choose corruptions i.i.d. from $\mathcal{N}(7,1)$ to simulate corruptions being large spikes in the signal (possibly caused by system malfunction or an adversarial agent). The maximum number of SRK iterations for each signal is $\tau=300$.  In Figure \ref{fig:f1a}, we create $\bPhi \in \mathbb{R}^{ m \times n} \overset{i.i.d}{\sim} \mathcal{N}(0,1)$ and $J = 300$. In Figure \ref{fig:f1b}, the entries of $\bPhi \in \mathbb{R}^{ m \times n }$ are drawn from a uniform distribution ranging from 0 to 1.
We note that, in both cases, Algorithm~\ref{alg:c_mmvsrk} is able to recover the full support after a sufficient number of iterations, whereas Algorithm~\ref{alg:mmvsrk} is only able to recover at most about 20\% of the support, regardless of the number of iterations. 
Since Algorithm~\ref{alg:mmvsrk} relies on row norms to estimate the joint support, it is to be expected that the relatively large value of the corruption would cause it to often be erroneously chosen to be part of the joint support estimate. 
As a result, this experiment highlights the advantage of the binning in Algorithm~\ref{alg:c_mmvsrk} in the presence of a single corruption with a high magnitude. 

\begin{figure}[ht]
\centering
\begin{subfigure}{0.4\textwidth}
\includegraphics[width=\textwidth]{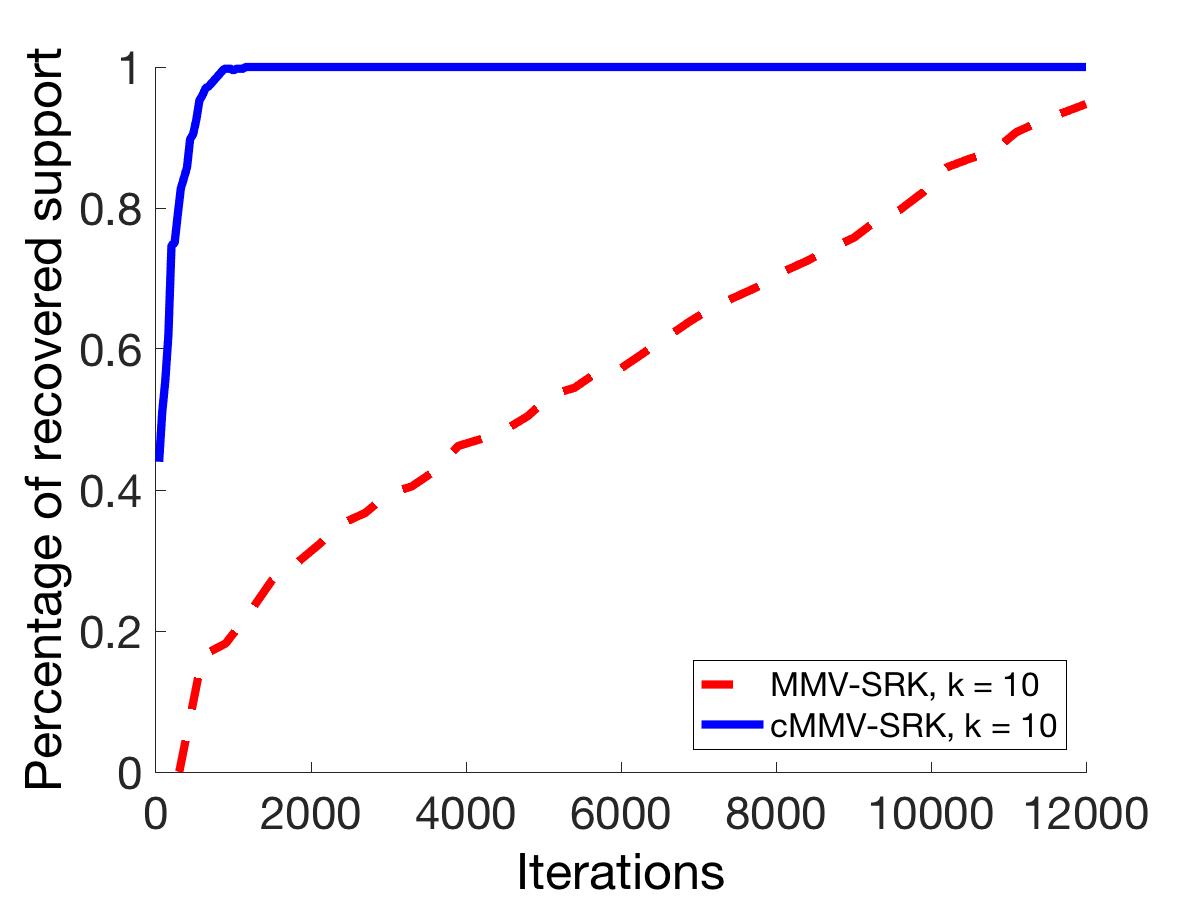}
\caption{$\bPhi \sim \mathcal{N}(0,1)$}
\label{fig:f2a}
\end{subfigure}
\begin{subfigure}{0.4\textwidth}
\includegraphics[width=\textwidth]{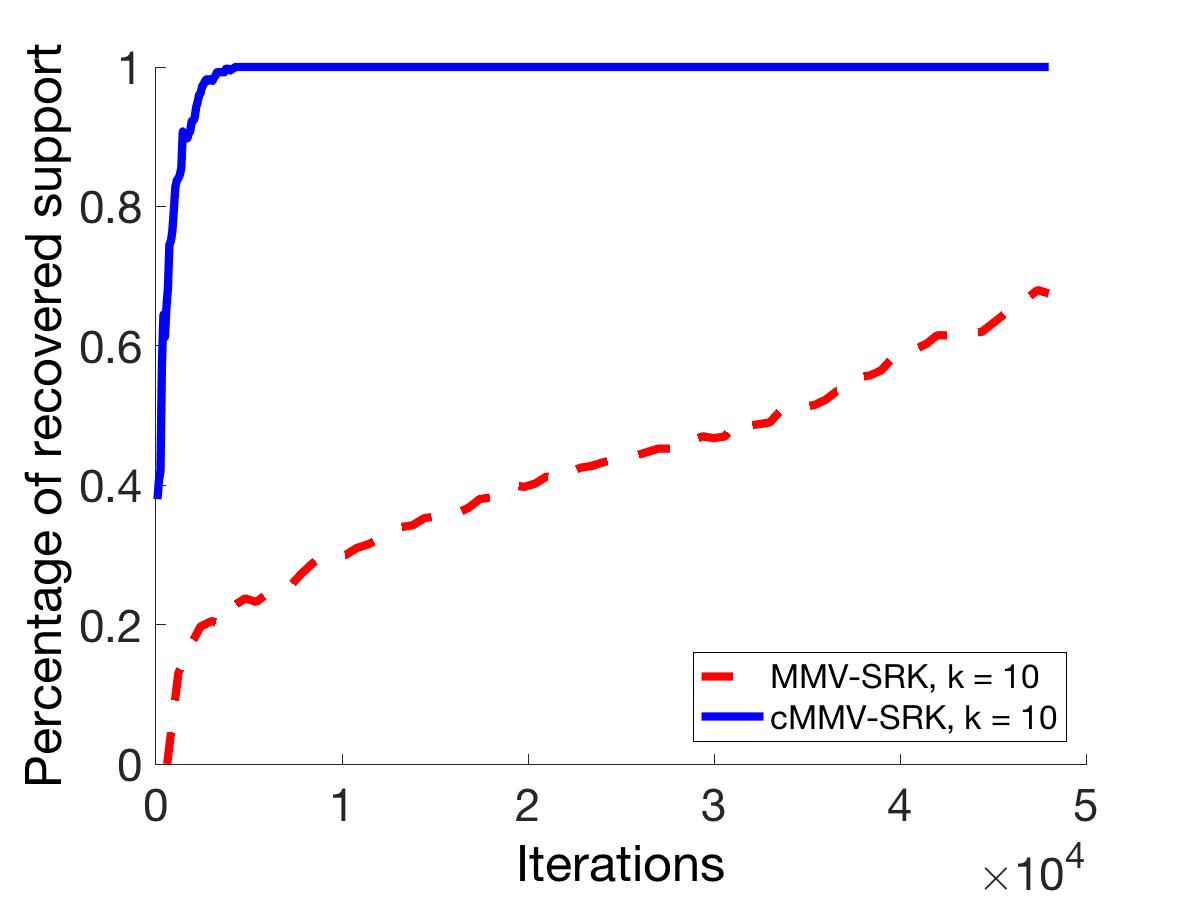}
\caption{$\bPhi \sim Unif([0,1])$}
\label{fig:f2b}
\end{subfigure}
\caption{Comparing SRK and MMV-SRK for support recovery when there is a single corrupt entry per signal whose magnitude is drawn from $\mathcal{N}(0,1)$.}
\label{fig:f2}
\end{figure}

In Figure \ref{fig:f2}, we experiment further with the magnitude of the corruption. Here we have that $m = 1000$, $n = 100$, and $k = 10$ but instead of the corrupt entries being drawn from a mean 7 and standard deviation 1 distribution, it is drawn from a standard normal distribution, as are the entries in the support. This allows us to test the robustness of our method to the choice of distribution. Note that Algorithm \ref{alg:mmvsrk} is able to find an increasingly accurate approximation for the support in this case, and will be able to recover the full support after a sufficiently large number of iterations. Because the magnitudes of the corruptions are smaller, the algorithm still has a chance of detecting the correct support using row norms to estimate $\jS$. However, Algorithm \ref{alg:c_mmvsrk} is able to obtain an accurate support estimate much faster than Algorithm \ref{alg:mmvsrk}.

\begin{figure}[h!]
\centering
\begin{subfigure}{0.4\textwidth}
\centering
\includegraphics[width=\textwidth]{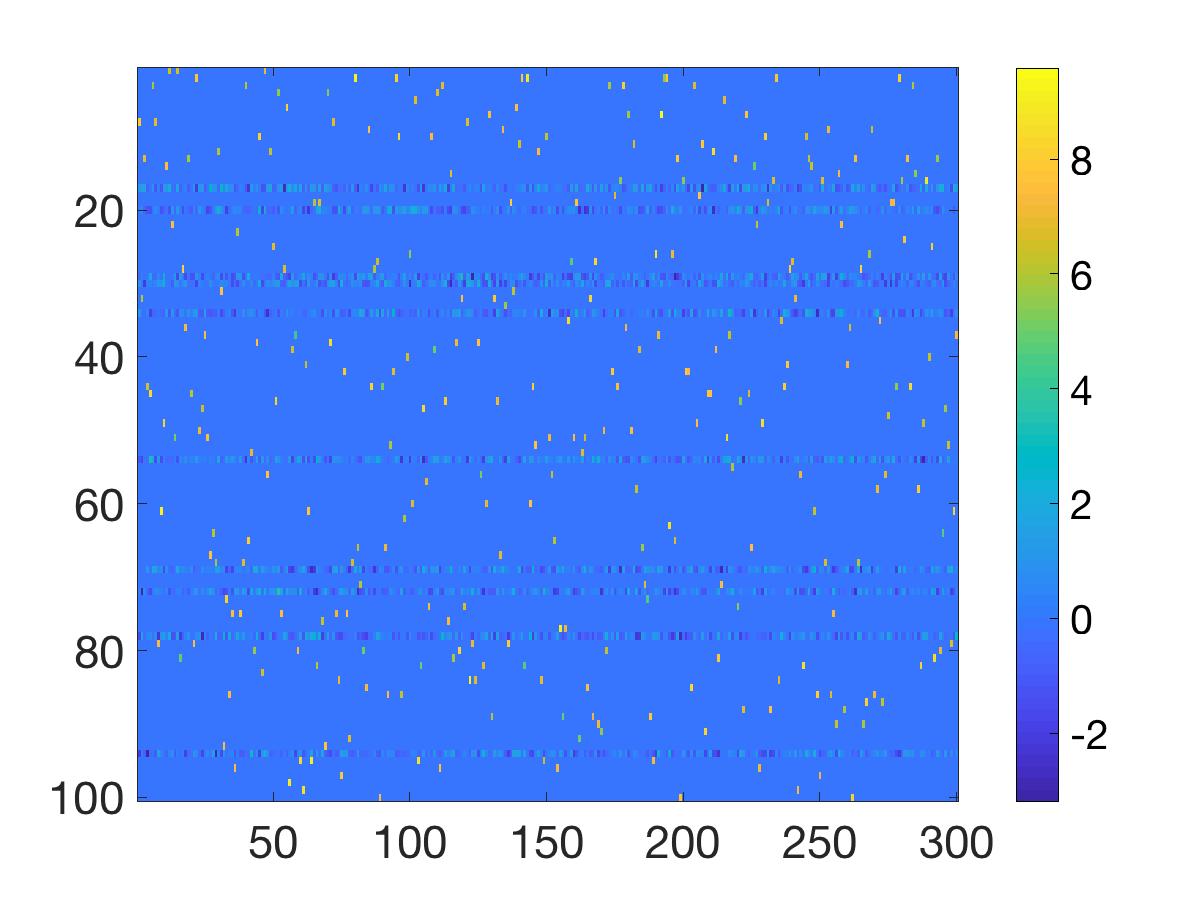}
\caption{$\X$ matrix with 1-3 corruptions per signal}
\label{fig:f3a}
\end{subfigure}
~
\begin{subfigure}{0.4\textwidth}
\centering
\includegraphics[width=\textwidth]{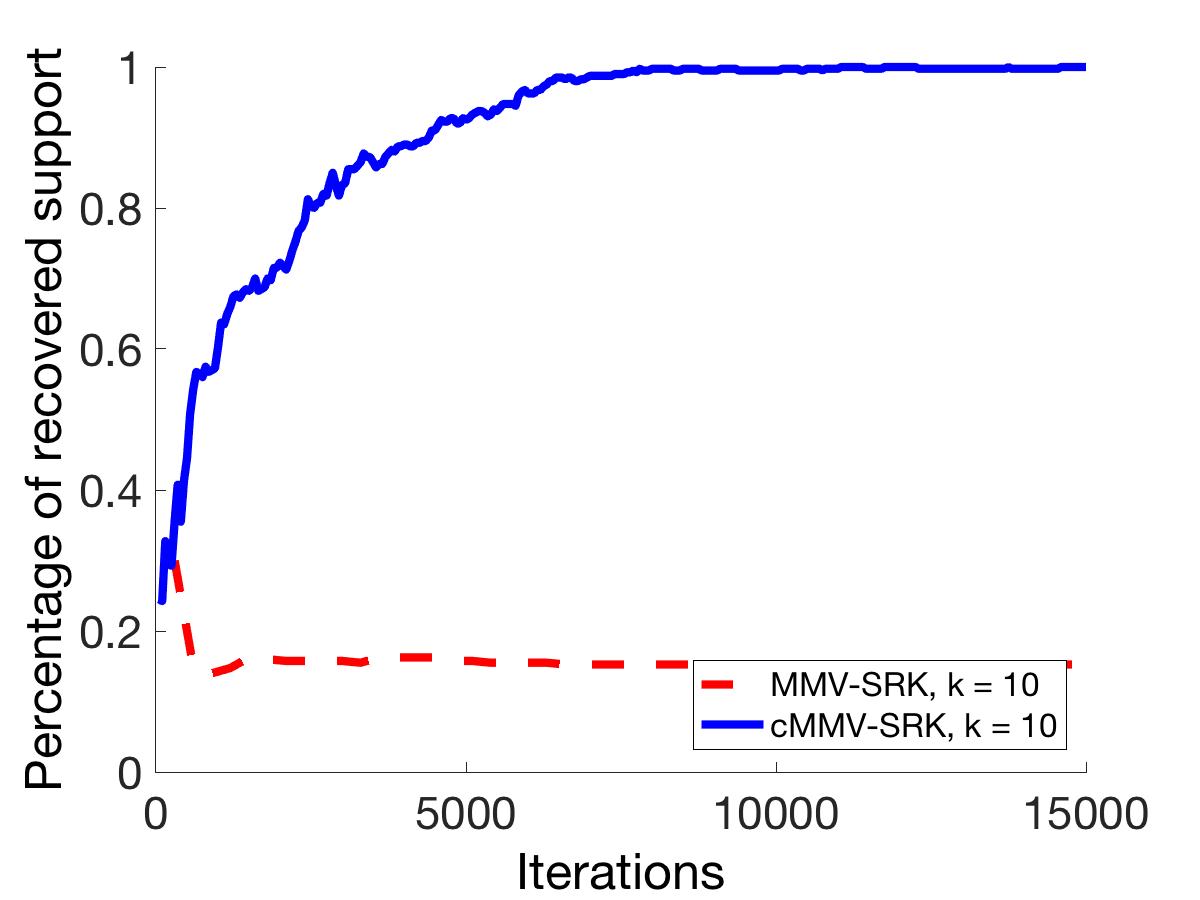}
\caption{Performance of Algorithm \ref{alg:c_mmvsrk} with multiple corruptions}
\label{fig:f3b}
\end{subfigure}
\caption{Investigating the robustness of cMMV-SRK when a random number (multiple) corruptions are introduced. Here, a signal can have between 1 and 3 corruptions. The corrupts magnitude of the corruptions are drawn from $\mathcal{N}(7,1)$.}
\label{fig:f3}
\end{figure}

In the next two experiments, we test the robustness of our proposed algorithm against multiple corruptions. In Figure \ref{fig:f3}, we allow for each signal to have multiple corruptions. For each signal, or column of $\X$, we uniformly at random select an integer from 1-3 to be the number of corruptions. The value of the corruptions are drawn i.i.d. from $\mathcal{N}(7,1)$ and an example of the resulting matrix can be seen in Figure \ref{fig:f3a}. The performance of the methods can be seen in Figure \ref{fig:f3b}.
We note that the results of this experiment are very similar to those of the experiment in Figure~\ref{fig:f1} since the corruptions are drawn from the same distribution. 
As we would expect, again due to the use of row norms, in the presence of multiple corruptions Algorithm~\ref{alg:mmvsrk} gives a less accurate estimate than in the presence of only one corruption drawn from this distribution, recovering no more than about 15\% of the support. 

\begin{figure}[ht]
\centering
\begin{subfigure}{.4\textwidth}
\includegraphics[width=\textwidth]{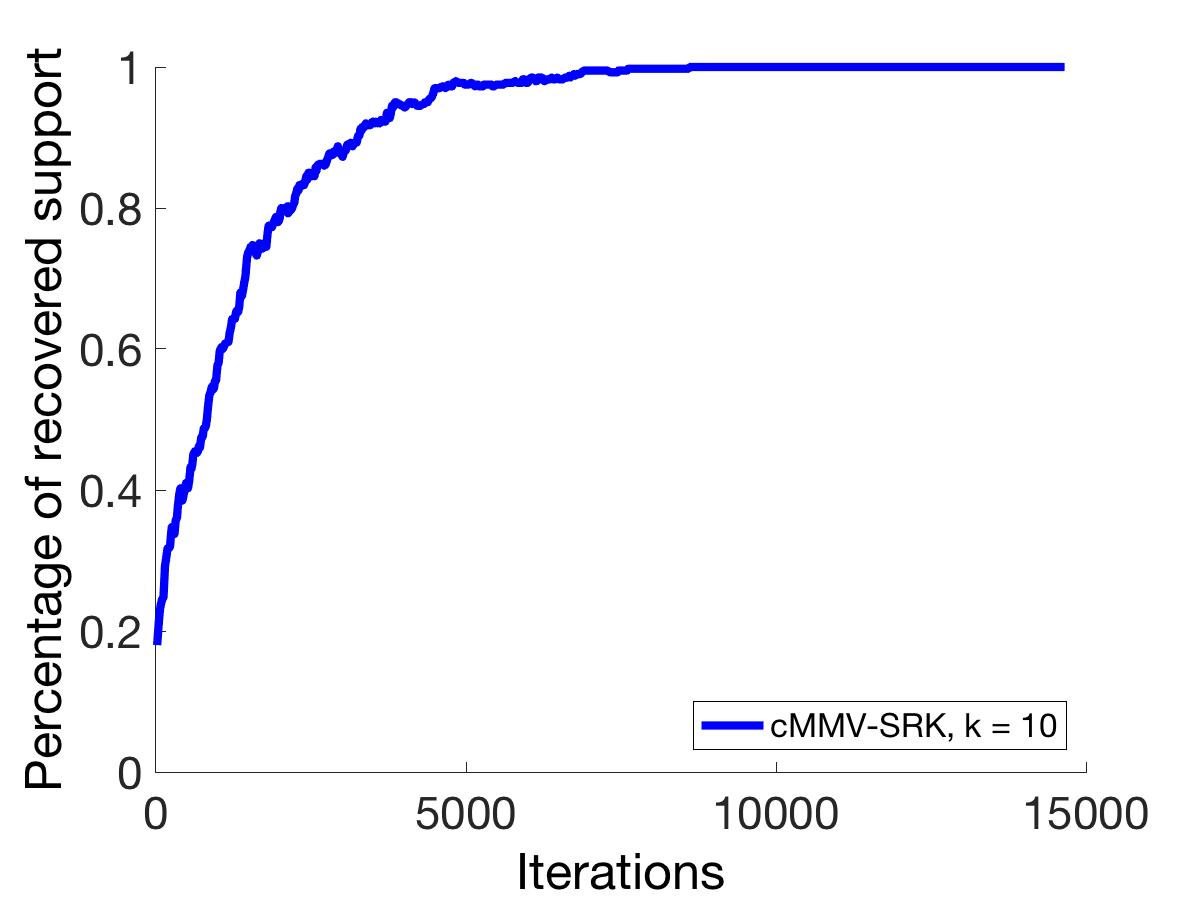}
\caption{Overdetermined linear systems}
\label{fig:f4}
\end{subfigure}
~
\begin{subfigure}{.4\textwidth}
\includegraphics[width=\textwidth]{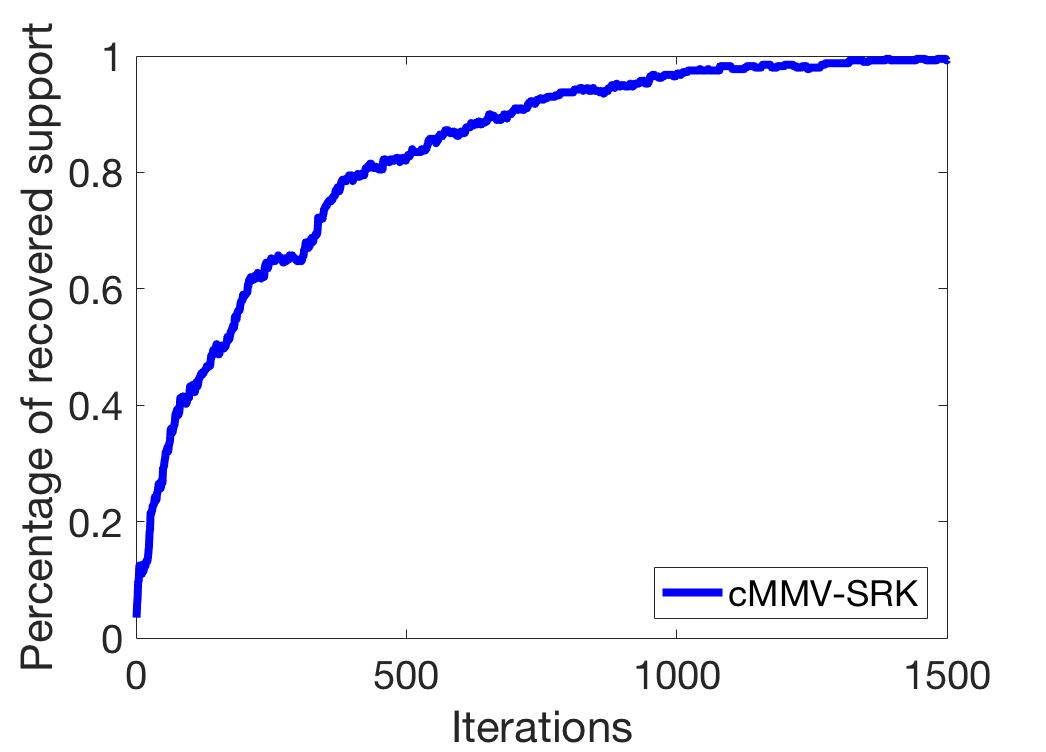}
\caption{Underdetermined linear systems.}
\label{fig:f5}
\end{subfigure}
\caption{Investigating the robustness of cMMV-SRK in a simulated online setting with a random number (multiple) corruptions. Here, a signal can have between 1 and 3 corruptions whose magnitudes are drawn from $\mathcal{N}(7,1)$ and we consider the over and under determined linear system settings.}
\label{fig:45}
\end{figure}

Figure \ref{fig:f4} shows the performance results for Algorithm \ref{alg:c_mmvsrk} in a simulated online setting. Instead of allowing the algorithm to loop for a fixed number of projections for each signal, we have 90\% of the signals with $\tilde{\tau}_j \in [5, 15]$ and the other 10\% of signals with $\tilde{\tau}_j \in [95, 100]$. The purpose of this is to simulate a variation in the amount of time a system has to work with a signal. The longer runs represent stalls in the online setting. For each signal, we first draw a random Bernoulli variable $z$ with probability of success $p = 0.1$. If $z=1$, then we choose an integer in $[95, 100]$ uniformly at random.  If $z=0$, then an integer in $[5, 15]$ is chosen uniformly at random. Algorithm \ref{alg:mmvsrk} cannot be investigated under this setting as the support estimate relies on processing all signals in every iteration.
We note that with respect to other parameters, the only difference between this experiment and that in Figure~\ref{fig:f3} is the size of $J$. We choose $J$ to be large enough such that the maximal number of projections made is 15000 (as in Figure~\ref{fig:f3}).

The following experiment is motivated by compressed sensing and utilizes an underdetermined linear systems as opposed to an overdetermined system. We repeat the the parameters as in Figure~\ref{fig:f4} with the exception of the measurement matrix, which has $m=100$ rows and $n = 500$ columns, and the total number of signals $J = 1500$. The results can be found in Figure~\ref{fig:f5}. In the underdetermined case, the proposed algorithm is still successful in recovering the support of the signal.

\begin{figure}[ht]
\centering
\begin{subfigure}{0.4\textwidth}
\centering
\includegraphics[width=.6\textwidth]{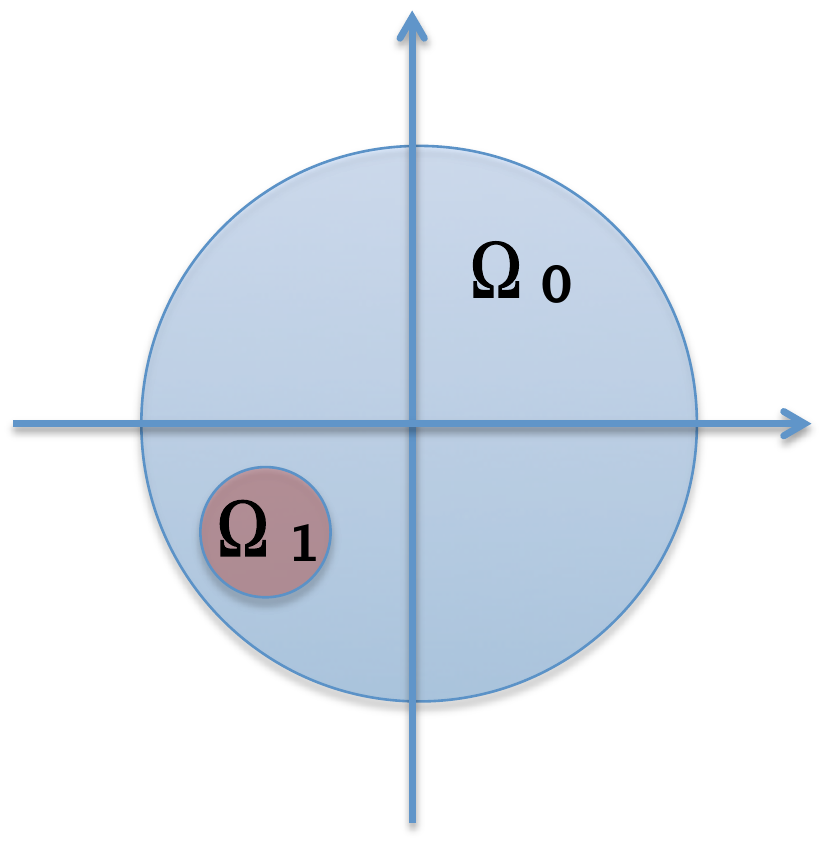}
\caption{Geometry of the real-world example.}
\label{fig:f6a}
\end{subfigure}
\begin{subfigure}{0.4\textwidth}
\includegraphics[width=1.1\textwidth]{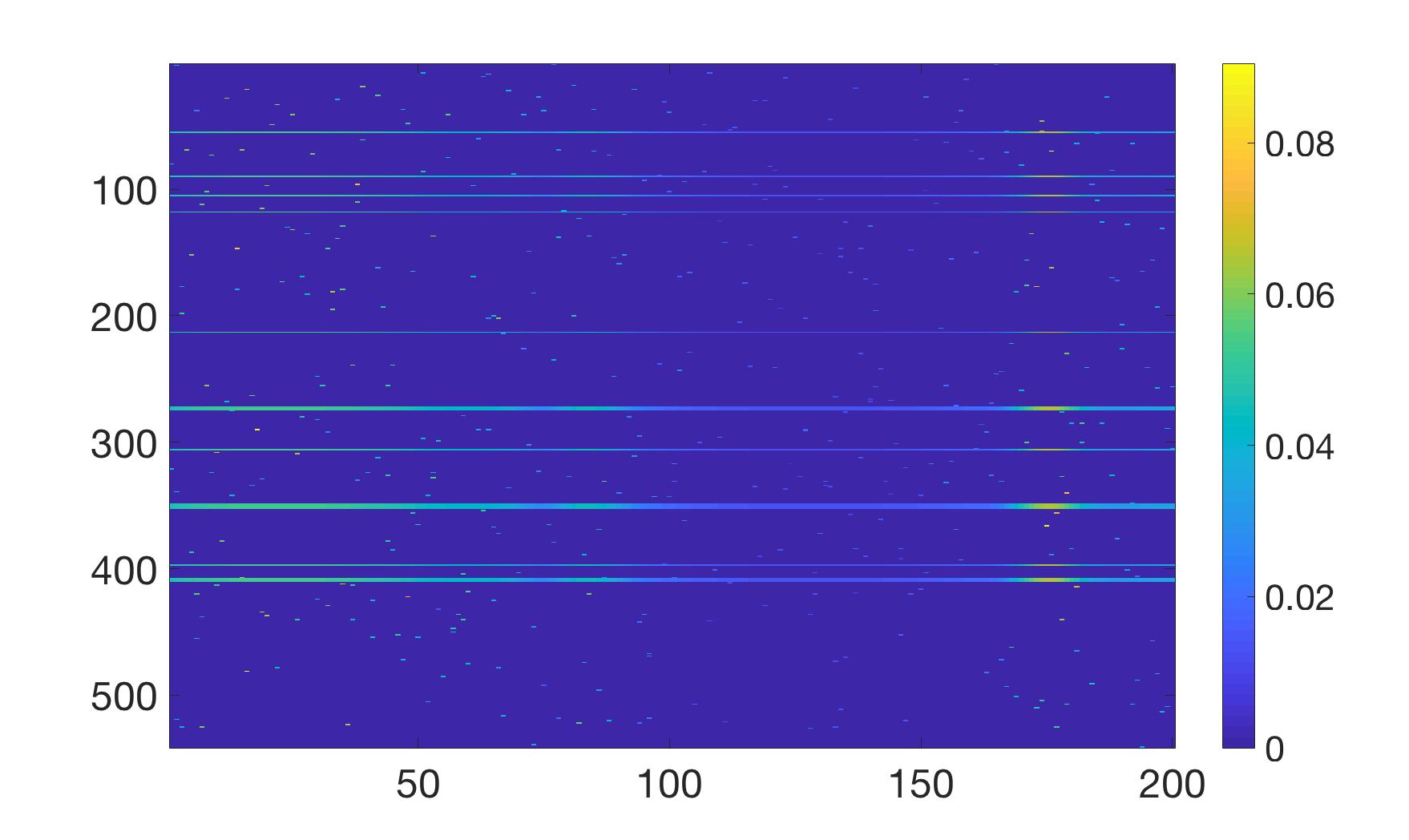}
\caption{$\mathbf{X}$ matrix with 1-3 corruptions per signal/wavelength}
\label{fig:f6b}
\end{subfigure}
\caption{Testing the robustness of cMMV-SRK in a real-world setting by using a simulated hyperspectral diffuse optical tomography (hyDOT) model. Healthy cells are in $\Omega_0$ while cancerous cells are located in $\Omega_1$. }
\label{fig:f6}
\end{figure}

\begin{figure}[ht]
\centering
\includegraphics[width=.5\textwidth]{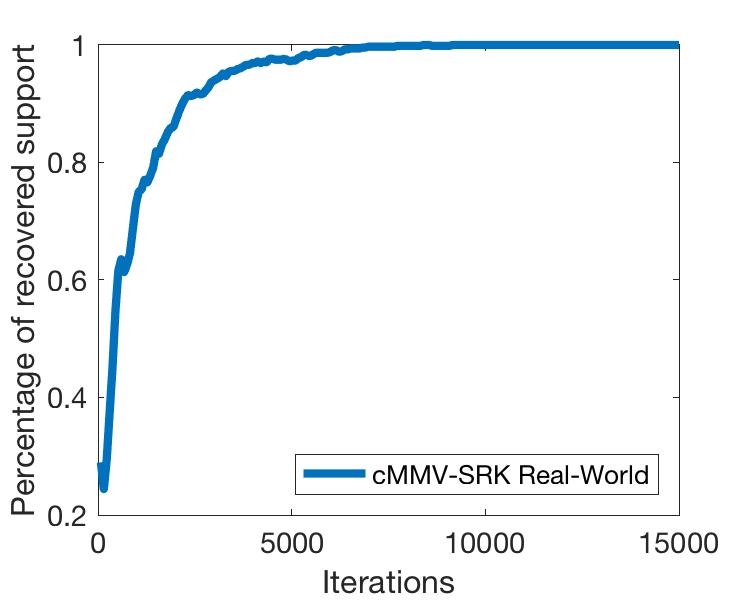}
\caption{Investigating the robustness of cMMV-SRK in a simulated real-world setting (hyperspectral diffuse optical tomography) when random multiple corruptions occur in each signal. In this setting, the measurement matrix is underdetermined ($m=248$, $n=541$).}
\label{fig:f7}
\end{figure}

Finally, we tested the robustness of our algorithm on the hyperspectral diffuse optical tomography motivating problem discussed in the introduction. For this experiment, we simulated absorption coefficient values for a two-dimensional circular sample of tissue of radius 25 centimeters, centered at the origin, with a circular tumor of radius 5 centimeters centered at the point (-15,-10). See Figure~\ref{fig:f6a}. Each signal was thus representing a reconstruction of the absorption coefficient value at each point in a mesh of size 541 over the sample area. The number of measurements corresponded to the number of source-detector pairs in the imaging process. We used a random Gaussian measurement matrix with $m = 248$ and $n = 541$, with $J = 200$ total signals, each corresponding to a different wavelength at which the tissue was imaged. We note that this is also an underdetermined system. As in previous experiments, 1 to 3 corruptions for each signal were drawn from a normal distribution with mean the average value of the absorption coefficient for the cancerous cells at each wavelength, and standard deviation a quarter of the distance between that value and the value of the absorption coefficient for the healthy cells. The online setting was not used for this experiment. The results can be found in Figure~\ref{fig:f7}. We see that the proposed algorithm is still successful in recovering the support of the signal. 

The experiments shown in this section highlights the usefulness of Algorithm \ref{alg:c_mmvsrk} for support recovery of jointly sparse MMVs, especially in the presence of (even quite large) corruptions. In each comparison between Algorithm \ref{alg:mmvsrk} and Algorithm \ref{alg:c_mmvsrk}, our proposed method outperforms the previously proposed method for support recovery. Additionally, the proposed method lends itself to the online setting. 

\section{Conclusion}
\label{sec:conclu}
In this work, we construct an algorithm for the support recovery of corrupted jointly sparse MMV. Our empirical results demonstrate that the proposed algorithm outperforms the previously proposed method, MMV-SRK, for support recovery of jointly sparse MMV, specifically when corruptions are present in the signal. Furthermore, empirical evidence indicates that our method is robust to the magnitude of corruptions and the number of corruptions. This improvement is due to the fact that the support estimate in Algorithm~\ref{alg:mmvsrk}, as many other signal recovery approaches for the jointly sparse MMV problem, depends on the row norms of the signals, which in this case would be dominated by the corruption In comparison, the estimate for Algorithm~\ref{alg:c_mmvsrk} only depends on the number of times an index appears in the support estimate of a signal. Lastly, our method lends itself well into the online setting when measurement vectors are streaming in continuously. We leave the analytical study of our method for future work.

\section*{Acknowledgements}
The initial  research for this  effort was conducted at the  Research Collaboration Workshop for Women in Data Science and Mathematics, July 17-21 held at ICERM. Funding for the workshop was provided by ICERM, AWM and DIMACS (NSF grant CCF-1144502). SL was supported by NSF CAREER grant CCF$-1149225$. DN was partially supported by the Alfred P. Sloan Foundation, NSF CAREER $\#1348721$, and NSF BIGDATA $\#1740325$. JQ was supported by the faculty start-up fund of Montana State University.

\bibliographystyle{abbrv}
\bibliography{../../../overall}

\end{document}